# Importance of local exact exchange potential in hybrid functionals for accurate excited states

*Jaewook Kim,[†] Kwangwoo Hong,[†] Sang-Yeon Hwang, Seongok Ryu, Sunghwan Choi, and Woo Youn Kim\**

Department of Chemistry, KAIST, 291 Daehak-ro, Yuseong-gu, Daejeon 34141, Republic of Korea

Density functional theory has been an essential analysis tool for both theoretical and experimental chemists since accurate hybrid functionals were developed. Here we propose a *local* hybrid method derived from the optimized effective potential (OEP) method and compare its distinct features with conventional *nonlocal* ones from the Hartree-Fock (HF) exchange operator. Both are formally exact for ground states and thus show similar accuracy for atomization energies and reaction barrier heights. For excited states, the local version yields virtual orbitals with $N$-electron character, while those of the nonlocal version have mixed characters between $N$- and $(N+1)$-electron orbitals. As a result, the orbital energy gaps from the former well approximate excitation energies with a small mean absolute error (MAE = 0.40 eV) for the Caricato benchmark set. The correction from time-dependent density functional theory with a simple local density approximation kernel further improves its accuracy by incorporating multi-configurational effects, resulting in the total MAE of 0.27 eV that outperforms conventional functionals except for MN15.

## 1. INTRODUCTION

Density functional theory (DFT) is apparently a predominant method for electronic structure calculations of molecules and solids. In particular, the emergence of hybrid functionals immediately attracted great attention in chemistry thanks to their reliable accuracy and versatile applicability to molecular systems.[1] As a result, DFT with hybrid functionals became a standard theoretical approach for various chemical applications such as elucidating chemical reactions at the atomistic level,[2,3] designing useful organic or inorganic materials,[4,5] and identifying spectroscopic data.[6]

The global hybrid functionals as the prototype is a simple combination between the exact exchange (EXX) energy ($E_{\text{EXX}}$) and (semi-)local exchange-correlation (xc) energy functionals ($E_{\text{xc}}^{\text{DFT}}$) as follows:

$$E_{\text{xc}}^{\text{hybrid}} = a_0 E_{\text{EXX}} + (1-a_0) E_{\text{x}}^{\text{DFT}} + E_{\text{c}}^{\text{DFT}}, \tag{1}$$

where the mixing ratio $a_0$ determines the portion of the Hartree-Fock (HF) exchange.[7,8] The fractional inclusion of the exact exchange partially cures the self-interaction error (SIE) in (semi-)local functionals that causes wrong ionization potentials (IPs) and electron affinities, upshifted highest occupied molecular orbital (HOMO) energies, small bandgap energies, etc.[9–13] Therefore, the mixing ratio tunes computational results for electronic properties of molecules. It is chosen either empirically[14] or formally.[8] B3LYP as an example of the former reproduces molecular geometries and binding energies to the similar accuracy of correlated ab initio methods but with much lower computational costs.[14,15] A new type of hybrid schemes is yet actively being developed to overcome their own limitations, e.g., meta-GGA,[16,17] double,[18,19] range-separated,[20–25] and local hybrid[26,27] functionals.

The hybrid methods are formally exact by virtue of the adiabatic connection theorem.[7,8,14,28] In practice, it can be solved within a single-determinant picture justified by the generalized Kohn-Sham (GKS) theory,[29]

$$\left[-\frac{1}{2}\nabla^2 + V_H(\mathbf{r}) + V_{ext}(\mathbf{r}) + a_0 V_{EXX}(\mathbf{r}) + (1-a_0)V_x^{DFT}(\mathbf{r}) + V_c^{DFT}(\mathbf{r})\right]\psi_i^{GKS}(\mathbf{r}) = \varepsilon_i^{GKS}\psi_i^{GKS}(\mathbf{r}), \quad (2)$$

where $V_H(\mathbf{r})$, $V_{ext}(\mathbf{r})$, $V_{x/c}^{DFT}(\mathbf{r})$, and $V_{EXX}(\mathbf{r})$ denote the Hartree, external, exchange/correlation potentials, and the exact exchange potential, respectively. Usually, $V_{EXX}(\mathbf{r})$ indicates the HF exchange operator,

$$V_{EXX}^{HF}(\mathbf{r}) = -\sum_j \hat{K}_j(\mathbf{r}), \quad \hat{K}_j(\mathbf{r})\psi_i^{GKS}(\mathbf{r}) = \left\{\int d\tau' \psi_j^{GKS*}(\mathbf{r}')\frac{1}{|\mathbf{r}-\mathbf{r}'|}\psi_i^{GKS}(\mathbf{r}')\right\}\psi_j^{GKS}(\mathbf{r}). \quad (3)$$

It is worthwhile to note that the exchange operator in eq 3 is nonlocal, so eq 2 does not correspond to the standard KS theory which requires the stringent locality of effective potentials. Alternatively, the exchange potential of the optimized effective potential (OEP)[30–32] ($V_{EXX}^{OEP}(\mathbf{r})$) derived from the HF exchange energy can be used. It also treats the exchange energy exactly and hence provides SIE-free DFT results and correct long-range behavior ($-1/r$ as $r \to \infty$).[33] Compared to the nonlocal exchange operator, the locality of OEP gives rise to distinct features as discussed in the literature.[34] First of all, both occupied and virtual orbitals feel the same effective potential made from the ground state density like conventional (semi-)local xc potentials. As a result, the OEP yields a number of bound virtual orbitals, whereas HF virtual orbitals are mostly unbound.[34–36] Furthermore, it has been known that the energy gaps between occupied and virtual orbitals of OEP approximate optical excitations including both valence and Rydberg transitions.[34,35,37] Compared to (semi-)local functionals, its correct asymptotic behavior gives HOMO energies close to the IPs, as the exact KS HOMO energy equals the exact IP.[38,39]

In this regard, it is interesting to study that to what extent a hybrid scheme with the EXX OEP is different from or similar to conventional ones and especially how those distinct features of OEP play a role in excited state calculations. Previous studies have shown similar or even better performance compared with conventional ones for nuclear magnetic resonance constants and band gap calculations.[40–45] To evaluate the general applicability of the OEP-based hybrid functionals to chemical problems, we implemented them in our KS-DFT code, namely ACE-Molecule,[35,36,46–49] and compared its computational results for general electronic properties of molecules with those of corresponding HF-based hybrid methods. We considered the PBE0 functional[8,50] but the distinct features coming from the locality of OEP can be applied to other hybrid functionals. For the sake of computational efficiency, we employed the Krieger–Li–Iafrate (KLI) approximation of OEP, $V_{\text{EXX}}^{\text{OEP}}(\mathbf{r}) \approx V_{\text{EXX}}^{\text{KLI}}(\mathbf{r})$ whose justification and expression can be found in the literature,[33,51–53] assuming that the main features of OEP due to its locality are also retained within the KLI approximation.

## 2. THEORY AND IMPLEMENTATION

**Derivation of local exchange-correlation potential from hybrid functional.** Here we derive a local xc potential from hybrid functionals using a similar perturbation theory that has been used in the derivation of the exchange-only KLI potential.[35] We assume that there is a non-interacting system whose ground-state density and energy are given by the GKS equation in eq 2:

$$\rho^{\text{GKS}}(\mathbf{r}) = \sum_{i}^{\text{occ}} \left| \psi_i^{\text{GKS}}(\mathbf{r}) \right|^2 \text{ and } E^{\text{GKS}}. \qquad (4)$$

Since eq 2 includes a non-local operator, one can derive a corresponding fully local KS equation using the OEP method that gives the same ground-state density and energy;

$$\hat{H}^{KS}\psi_i^{KS}(\mathbf{r}) = \left[-\frac{1}{2}\nabla^2 + V_H(\mathbf{r}) + V_{ext}(\mathbf{r}) + v_{xc}^{OEP}(\mathbf{r})\right]\psi_i^{KS}(\mathbf{r}) = \varepsilon_i^{KS}\psi_i^{KS}(\mathbf{r}), \quad (5)$$

and

$$\rho^{KS}(\mathbf{r}) = \sum_i^{occ}\left|\psi_i^{KS}(\mathbf{r})\right|^2 = \rho^{GKS}(\mathbf{r}) \text{ and } E^{KS} = E^{GKS}, \quad (6)$$

where $v_{xc}^{OEP}(\mathbf{r})$ denotes an OEP xc potential. Eq 5 with an aribitrary OEP xc potential may produce different ground-state density and energy from those of the corresponding GKS equation. Then, we introduce a perturbative correction term, $\Delta\hat{v}$, in the KS equation to minimize their energy and density differences. The perturbative term in this case is defined by the difference between the KS and GKS Hamiltonians,

$$\Delta\hat{v} = \hat{H}^{GKS} - \hat{H}^{KS} = -a_0\sum_j \hat{K}_j(\mathbf{r}) + (1-a_0)V_x^{DFT}(\mathbf{r}) + V_c^{DFT}(\mathbf{r}) - v_{xc}^{OEP}(\mathbf{r}). \quad (7)$$

Considering only up to the first order correction, the modified KS equation and its density become

$$\left(\hat{H}^{KS} + \eta\Delta\hat{v}\right)\left(\psi_i^{KS} + \eta\psi_i^{(1)} + \mathcal{O}(\eta^2)\right) = \left(\varepsilon_i^{KS} + \eta\varepsilon_i^{(1)} + \mathcal{O}(\eta^2)\right)\left(\psi_i^{KS} + \eta\psi_i^{(1)} + \mathcal{O}(\eta^2)\right) \quad (8)$$

and

$$\begin{aligned}\rho^{KS}(\mathbf{r}) &= \sum_i^{occ}\left|\psi_i^{KS}(\mathbf{r}) + \eta\psi_i^{(1)}(\mathbf{r}) + \mathcal{O}(\eta^2)\right|^2 \\ &= \sum_i^{occ}\left\{\left|\psi_i^{KS}(\mathbf{r})\right|^2 + \left(\eta\psi_i^{(1)*}(\mathbf{r})\psi_i^{KS}(\mathbf{r}) + \text{c.c.}\right)\right\} + \mathcal{O}(\eta^2),\end{aligned} \quad (9)$$

respectively, where $\eta$ denotes an perturbation order parameter, and $\varepsilon_i^{(1)}$ and $\psi_i^{(1)}$ are the first-order corrections of $\varepsilon_i^{KS}$ and $\psi_i^{KS}$, respectively. Using eq 7, $\varepsilon_i^{(1)}$ can be written as

$$\begin{aligned}\varepsilon_i^{(1)} &= \langle \psi_i^{KS} | \Delta \hat{v} | \psi_i^{KS} \rangle \\ &= \langle \psi_i^{KS} | -a_0 \sum_j \hat{K}_j(\mathbf{r}) + (1-a_0)V_x^{DFT}(\mathbf{r}) + V_c^{DFT}(\mathbf{r}) - v_{xc}^{OEP}(\mathbf{r}) | \psi_i^{KS} \rangle \\ &\equiv a_0 \overline{v}_{x,i} + (1-a_0)\overline{V}_{x,i}^{DFT} + V_{c,i}^{DFT} - \overline{v}_{xc,i}^{OEP}\end{aligned} \quad (10)$$

and $\psi_i^{(1)}$ is given by

$$\begin{aligned}\psi_i^{(1)}(\mathbf{r}) &= \sum_{k=1, k\neq i}^{\infty} \frac{\langle \psi_k^{KS} | \Delta \hat{v} | \psi_i^{KS} \rangle}{\varepsilon_k - \varepsilon_i} \psi_k^{KS}(\mathbf{r}) \\ &= \sum_{k=1, k\neq i}^{\infty} \frac{\langle \psi_k^{KS} | -a_0 \sum_j \hat{K}_j(\mathbf{r}) + (1-a_0)V_x^{DFT}(\mathbf{r}) + V_c^{DFT}(\mathbf{r}) - v_{xc}^{OEP}(\mathbf{r}) | \psi_i^{KS} \rangle}{\varepsilon_k - \varepsilon_i} \psi_k^{KS}(\mathbf{r}).\end{aligned} \quad (11)$$

Substituting eq 9 into eq 6, the following OEP integral equation can be obtained.

$$\begin{aligned}0 &= \rho^{KS}(\mathbf{r}) - \rho^{GKS}(\mathbf{r}) \\ &= \sum_i^{occ} \left\{ |\psi_i^{KS}(\mathbf{r})|^2 + \left( \eta \psi_i^{(1)*}(\mathbf{r}) \psi_i^{KS}(\mathbf{r}) + c.c. \right) \right\} - \sum_i^{occ} |\psi_i^{GKS}(\mathbf{r})|^2 + \mathcal{O}(\eta^2) \\ &= \eta \sum_i^{occ} \left( \psi_i^{(1)*}(\mathbf{r}) \psi_i^{KS}(\mathbf{r}) + c.c. \right) + \mathcal{O}(\eta^2)\end{aligned} \quad (12)$$

To obtain the expression of $v_{xc}^{OEP}(\mathbf{r})$, eq 8 can be expanded as

$$\hat{H}^{KS} \psi_i^{KS} + \eta \Delta \hat{v} \psi_i^{KS} + \eta \hat{H}^{KS} \psi_i^{(1)} = \varepsilon_i^{KS} \psi_i^{KS} + \eta \varepsilon_i^{KS} \psi_i^{(1)} + \eta \varepsilon_i^{(1)} \psi_i^{KS} + \mathcal{O}(\eta^2). \quad (13)$$

Then, the first-order terms lead to the following equation,

$$\begin{aligned}\left( \Delta \hat{v} - \varepsilon_i^{(1)} \right) \psi_i^{KS} &= \left( \varepsilon_i^{KS} - \hat{H}^{KS} \right) \psi_i^{(1)} \\ &= \left( \varepsilon_i^{KS} + \frac{1}{2} \nabla^2 - V_H(\mathbf{r}) - V_{ext}(\mathbf{r}) - v_{xc}^{OEP}(\mathbf{r}) \right) \psi_i^{(1)}.\end{aligned} \quad (14)$$

Multiplying $V_H(\mathbf{r}) + V_{ext}(\mathbf{r}) + v_{xc}^{OEP}(\mathbf{r})$ to eq 12 and using eq 14, it gives

$$0 = \sum_i^{occ} \left\{ \left( \varepsilon_i^{KS} + \frac{1}{2}\nabla^2 \right) \psi_i^{(1)*}(\mathbf{r}) - \left( \Delta\hat{v} - \varepsilon_i^{(1)} \right) \psi_i^{KS*}(\mathbf{r}) \right\} \psi_i^{KS}(\mathbf{r}) + c.c.. \tag{15}$$

By using eqs 7 and 10, eq 15 can be rearranged to obtain

$$\begin{aligned}v_{xc}^{OEP}(\mathbf{r}) = \frac{1}{\rho_0(\mathbf{r})} \sum_i^{occ} &\left[ \left( \varepsilon_i^{KS} + \frac{1}{2}\nabla^2 \right) \psi_i^{(1)*}(\mathbf{r}) \psi_i^{KS}(\mathbf{r}) \right.\\ &+ \psi_i^{KS*}(\mathbf{r}) \left\{ -a_0 \sum_j \hat{K}_j(\mathbf{r}) + (1-a_0)V_x^{DFT}(\mathbf{r}) + V_c^{DFT}(\mathbf{r}) \right.\\ &\left.\left. - \left( a_0 \bar{v}_{x,i} + (1-a_0)\bar{V}_{x,i}^{DFT} + \bar{V}_{c,i}^{DFT} - \bar{v}_{xc,i}^{OEP} \right) \right\} \psi_i^{KS}(\mathbf{r}) \right] + c.c..\end{aligned} \tag{16}$$

Finally, the KLI approximation to eq 16 that is

$$\langle \psi_k^{KS} | \Delta\hat{v} | \psi_i^{KS} \rangle = 0 \quad \text{for all } k \neq i \tag{17}$$

results in $\psi_i^{(1)}(\mathbf{r}) = 0$ and thus

$$\begin{aligned}v_{xc}^{OEP}(\mathbf{r}) \approx \frac{1}{\rho^{KS}(\mathbf{r})} \sum_i^{occ} &\left[ \psi_i^{KS*}(\mathbf{r}) \left\{ -a_0 \sum_j \hat{K}_j(\mathbf{r}) + (1-a_0)V_x^{DFT}(\mathbf{r}) + V_c^{DFT}(\mathbf{r}) \right.\right.\\ &\left.\left. - \left( a_0 \bar{v}_{x,i} + (1-a_0)\bar{V}_{x,i}^{DFT} + \bar{V}_{c,i}^{DFT} - \bar{v}_{xc,i}^{OEP} \right) \right\} \psi_i^{KS}(\mathbf{r}) \right] + c.c..\end{aligned} \tag{18}$$

Referring to the expression of the exchange-only KLI potential,

$$v_x^{KLI}(\mathbf{r}) = \frac{1}{\rho^{KS}(\mathbf{r})} \sum_i^{occ} \left[ \psi_i^{KS*}(\mathbf{r}) \left\{ -\sum_j \hat{K}_j(\mathbf{r}) - \left( \bar{v}_{x,i} - \bar{v}_{x,i}^{KLI} \right) \right\} \psi_i^{KS}(\mathbf{r}) \right] + c.c., \tag{19}$$

the so-called KLI hybrid potential, $v_{xc}^{KLI}(\mathbf{r})$, can be represented as

$$v_{xc}^{KLI}(\mathbf{r}) = a_0 v_x^{KLI}(\mathbf{r}) + (1-a_0)V_x^{DFT}(\mathbf{r}) + V_c^{DFT}(\mathbf{r}). \tag{20}$$

Finally, we arrive at the following KS equation with the KLI hybrid potential that is supposed to give the same ground-state density and energy with conventional HF-based hybrid functionals,

$$\left[-\frac{1}{2}\nabla^2 + V_H(\mathbf{r}) + V_{ext}(\mathbf{r}) + v_{xc}^{KLI}(\mathbf{r})\right]\psi_i(\mathbf{r}) = \varepsilon_i \psi_i(\mathbf{r}). \tag{21}$$

The only difference between the GKS equation for hybrid functionals and its corresponding KS equation with the KLI approximation is that the HF exchange operator is replaced by the exchange-only KLI potential. Hereafter, PBE0KLI-$a_0$ and PBE0HF-$a_0$ denote KLI- and HF-based hybrid methods with the exact exchange portion $a_0$ in the form of eq 1, respectively.

**Implementation of KLI-based hybrid method.** We implemented the KLI-based hybrid method of eq 20 in our ACE-Molecule program. To perform the integral in eqs 3 and 19, we adopted the interpolating scaling function method, [54–59]

$$\int d\tau' \phi_j^*(\mathbf{r}') \frac{1}{|\mathbf{r}-\mathbf{r}'|} \phi_i(\mathbf{r}') = \frac{2}{\sqrt{\pi}} \int_0^\infty dt \int d\tau' \phi_j^*(\mathbf{r}') \phi_i(\mathbf{r}') e^{-t^2(\mathbf{r}-\mathbf{r}')^2}$$
$$= \frac{2}{\sqrt{\pi}} \sum_a w_a \sum_i F_{i'i}^{x,a} \sum_j F_{j'j}^{y,a} \sum_k F_{k'k}^{z,a} d_{ijk} + \frac{\pi}{t_f^2} \phi_j^*(\mathbf{r}) \phi_i(\mathbf{r}), \tag{22}$$

where $F_{i'i}^{x,a}$, $F_{j'j}^{y,a}$, and $F_{k'k}^{z,a}$ are defined as

$$F_{i'i}^{x,a} = \int_{-\infty}^{\infty} e^{-t_a^2(x_{i'}-x)^2} L_i(x) dx, \tag{23}$$

and $w_a$ and $d_{ijk}$ are a weight factor and a expansion coefficient of $\phi_j^*(\mathbf{r})\phi_i(\mathbf{r})$ with the product of Lagrange-sinc functions, $L_i(x)L_j(y)L_k(z)$, respectively. The remaining terms in eq 19 can be evaluated by the standard procedure as implemented in Refs. 35 and 60. To compute the remaining terms in the KLI hybrid potential, eq 20, the Libxc library[61] was used.

## 3. COMPUTATIONAL DETAILS

Gaussian 09[62] and ACE-Molecule[35,36,46–49] were used for all the HF and KLI hybrid calculations, respectively. For the ACE-Molecule program, the spherical simulation box was adopted except for excited state benchmark calculations with namely the Caricato set[63] for which we constructed a simulation box by combining the spheres around each atom with the radius of 15.1 Bohr. In ACE-Molecule, the pseudopotential method was adopted to describe nuclear potentials. We used the pseudopotentials for PBE in the standard solid-state pseudopotentials library by Dal Corso[64]. The geometry of molecules for atomization energies and barrier heights were obtained from Ref. 65. The Cartesian coordinate of ground-state thiophenol is given in Table S1, and those of the molecules in the Caricato set are given in Table S2-S12. The supersampling method[47] was used to suppress the egg-box effect in grid-based calculations. Table 1 summarizes the detailed computational parameters used in this work. To express Rydberg states correctly, we adopted the d-aug-cc-pVDZ basis set[66] for Caricato set calculations. The CASSCF and CASPT2 calculations of thiophenol were performed with the (12,11) active space and the 6-311++G(d,p) basis set.

**Table 1.** Basis sets for HF hybrid calculations, and the radius of simulation box and the grid spacing for KLI hybrid calculations

|  | HF hybrid | KLI hybrid | |
|---|---|---|---|
|  | Basis set | Radius (Bohr) | Grid spacing (Bohr) |
| Atomization energy | aug-cc-pV5Z[a] | 12.5 | 0.156 |
| Barrier height | aug-cc-pV5Z | 10.0 | 0.167 |
| Caricato set | d-aug-cc-pVDZ | 15.1[b] | 0.300 |

| Potential energy curve of Thiophenol | cc-pVTZ | 12.0 | 0.300 |

[a] For SiH$_4$, the aug-cc-pVQZ basis set was used due to the convergence problem.

[b] Superposition of spheres with the radius of 15.1 Bohr around each atom.

## 4. RESULTS AND DISCUSSION

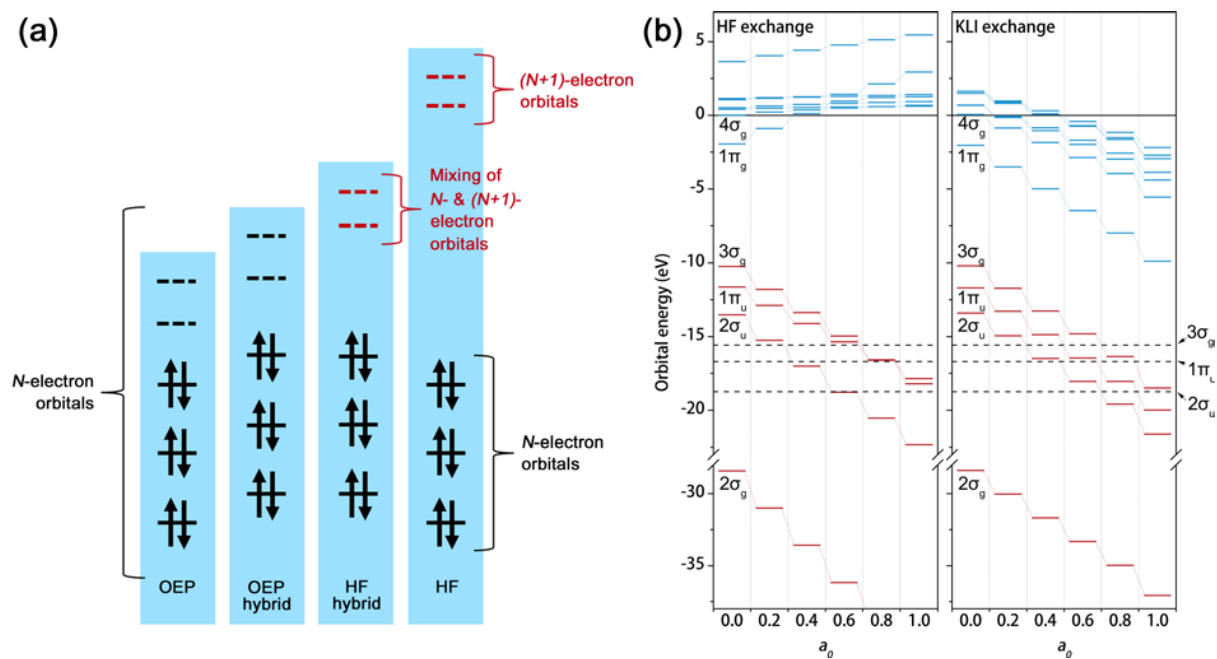

**Figure 1.** (a) Conceptual representation of OEP and HF hybrid orbitals. (b) The change in the orbital energies of N$_2$ as a function of the mixing ratio $a_0$. The red and blue lines denote occupied and virtual orbitals, respectively. The dashed lines indicate the experimental ionization potentials obtained from Ref. 67.

**Conceptual differences between OEP and HF hybrid methods.** Figure 1a conceptually shows the key difference between the OEP and HF hybrid methods. An exchange-only OEP potential is designed to give the corresponding HF density with $N$ electrons. Thus, occupied orbitals from both methods are similar and correspond to $N$-electron orbitals. However, HF virtual orbitals correspond to $(N+1)$-electron orbitals, while the exchange-only OEP (also KS) gives $N$-electron virtual orbitals.[13,68] Therefore, the energies of HF virtual orbitals are much higher (usually above the vacuum level) than those of OEP. In this context, the occupied and virtual orbitals of the HF or OEP hybrids are affected differently by the mixing ratio of exchange term. Both hybrids yield $N$-electron occupied orbitals regardless of the mixing ratio. However, the former gives rise to virtual orbitals with characters mixed between $N$- and $(N+1)$-electrons depending on the mixing ratio, whereas the latter always produces $N$-electron ones. As a result, the energy gaps between the occupied and virtual orbitals of HF hybrid strongly depend on the mixing ratio; the higher the ratio, the larger the energy gaps, because of the stronger repulsion of an electron in the virtual orbitals.

Figure 1b, which compares the orbital energies of $N_2$ between PBE0KLI-$a_0$ and PBE0HF-$a_0$ as a function of the mixing ratio, indeed verifies the above perception. The occupied orbitals from both potentials show a similar trend with one another; their energies gradually decrease, as the mixing ratio increases, due to the partial alleviation of SIE in the PBE[69,70] functional. Futhermore, PBE0HF-$a_0$ functional provides HOMO with wrong symmetry ($1\pi_u$) in the caes of high mixing ratio, which is due to the well known problem of the HF method.[71] At some high mixing ratio, their occupied orbital energies become close to the IPs of $N_2$ indicated by the dashed lines, which reflects the Koopman's theorem of DFT.[72–74] However, their virtual orbitals behave in a different way due to the different nature of the HF and OEP exchange terms. As the mxing ratio increases, the energy gaps between the occupied and virtual orbitals of HF hybrid increase, whereas those of the KLI hybrid remain almost constant.

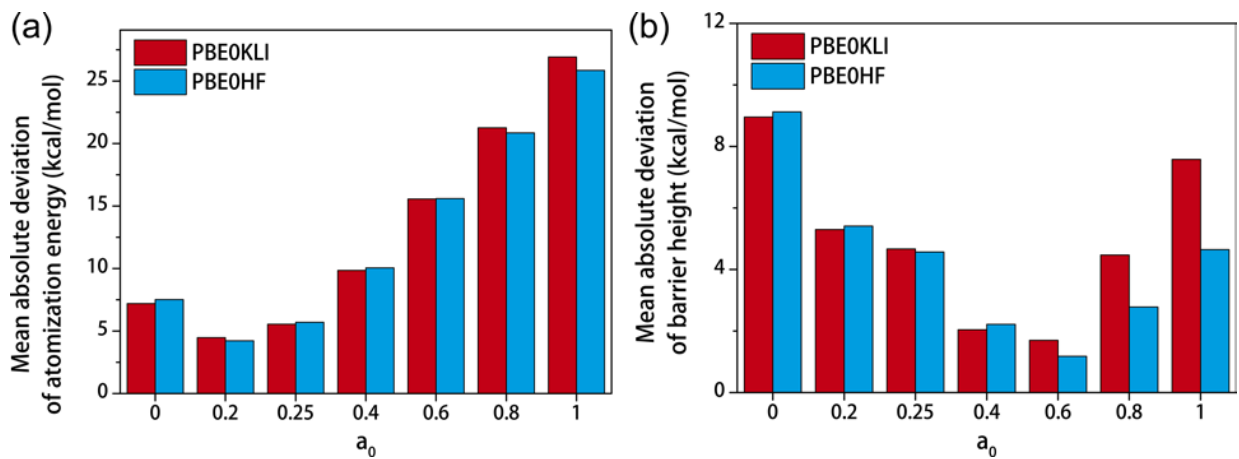

**Figure 2.** The mean absolute deviation of (a) the atomization energies of $P_2$, HF, $F_2$, and $SiH_4$ and (b) the kinetic energy barriers of the BH6 data set[75] with respect to those obtained from ab initio calculations.[75,76] $a_0$ is the mixing ratio of exact exchange.

**Difference between OEP and HF hybrid methods: ground states.** Subsequently, we investigate how such differences affect ground- and excited-state properties of molecular systems. For ground states, atomization energies and kinetic energy barriers were considered. Figure 2 shows the mean absolute deviation (MAD) of the atomization energies of $P_2$, HF, $F_2$, and $SiH_4$ and the kinetic energy barriers of the BH6 data set[75] with respect to highly accurate computational results[75,76] as a function of $a_0$. The KLI and HF hybrids show similar trends as expected and minimize the MAD near their original $a_0$ value (0.25). Individual values for each molecule are also very close to one another (Figure S1). For both KLI and HF hybrids, the MAD values of the kinetic energy barriers are also similar. Such similarity for the ground-state properties seems natural because they are determined by the occupied orbitals.

**Difference between OEP and HF hybrid methods: excited states.** To compare between the HF and OEP hybrids for excited states, we calculated 30 valence and 39 Rydberg excitation energies of the following 11 molecules in the Caricato set[63]: formaldehyde, acetaldehyde, acetone, ethylene, isobutene, trans-butadiene, pyridine, pyrazine, pyridazine, pyrimidine, and s-tetrazine. Table 2 shows the orbital energy gaps from PBE0KLI-$a_0$ with various $a_0$ as an approximation to each excitation energy and the corresponding experimental values.

**Table 2.** The orbital energy gaps from PBE0KLI-$a_0$ with various $a_0$ as an approximation to each excitation energy for the Caricato set. The third column indicates the type of excitations: V and R for valence and Rydberg excitations, respectively. The experimental values were obtained from Ref. 63. Unit: eV

| Molecule | State | Type | PBE0KLI-a0 (Orbital energy gaps) | | | | | | | Expt. |
|---|---|---|---|---|---|---|---|---|---|---|
| | | | 0.0 | 0.2 | 0.4 | 0.5 | 0.6 | 0.8 | 1.0 | |
| formaldehyde | $1^1A_2$ | V | 3.46 | 3.59 | 3.71 | 3.77 | 3.83 | 3.95 | 4.28 | 4.00 |
| | $1^1B_2$ | R | 5.83 | 6.41 | 6.96 | 7.22 | 7.48 | 7.98 | 8.49 | 7.08 |
| | $2^1B_2$ | R | 6.67 | 7.23 | 7.76 | 8.00 | 8.24 | 8.69 | 9.17 | 7.97 |
| | $2^1A_1$ | R | 6.61 | 7.28 | 7.92 | 8.22 | 8.52 | 9.10 | 9.75 | 8.14 |
| | $2^1A_2$ | R | 6.92 | 7.53 | 8.13 | 8.42 | 8.70 | 9.27 | 10.00 | 8.37 |
| | $3^1B_2$ | R | 7.46 | 8.13 | 8.75 | 9.05 | 9.34 | 9.88 | 10.55 | 8.88 |
| | $1^1B_1$ | V | 8.07 | 8.24 | 8.40 | 8.48 | 8.56 | 8.71 | 9.00 | 9.00 |
| | $3^1A_2$ | R | 7.73 | 8.44 | 9.14 | 9.50 | 9.84 | 10.54 | 11.52 | 9.22 |
| | $4^1B_2$ | R | 7.66 | 8.36 | 9.05 | 9.39 | 9.73 | 10.39 | 11.27 | 9.26 |
| | $4^1A_1$ | R | 8.23 | 8.94 | 9.66 | 10.02 | 10.38 | 11.10 | 12.20 | 9.58 |
| | $5^1B_2$ | R | 7.71 | 8.46 | 9.21 | 9.59 | 9.97 | 10.73 | 11.86 | 9.63 |
| acetaldehyde | $1^1A''$ | V | 3.84 | 3.99 | 4.14 | 4.21 | 4.28 | 4.42 | 4.75 | 4.28 |
| | $2^1A'$ | R | 5.38 | 5.92 | 6.42 | 6.66 | 6.90 | 7.34 | 7.78 | 6.82 |
| | $3^1A'$ | R | 5.94 | 6.55 | 7.12 | 7.39 | 7.66 | 8.16 | 8.66 | 7.46 |
| | $4^1A'$ | R | 6.33 | 6.96 | 7.55 | 7.82 | 8.09 | 8.60 | 9.10 | 7.75 |
| | $6^1A'$ | R | 6.90 | 7.56 | 8.19 | 8.50 | 8.79 | 9.35 | 9.98 | 8.43 |
| | $7^1A'$ | R | 7.07 | 7.78 | 8.47 | 8.80 | 9.13 | 9.76 | 10.51 | 8.69 |
| acetone | $1^1A_2$ | V | 3.93 | 4.11 | 4.27 | 4.35 | 4.44 | 4.59 | 4.91 | 4.43 |
| | $1^1B_2$ | R | 4.96 | 5.49 | 5.98 | 6.21 | 6.44 | 6.87 | 7.27 | 6.36 |
| | $2^1A_2$ | R | 5.93 | 6.56 | 7.16 | 7.44 | 7.72 | 8.24 | 8.77 | 7.36 |
| | $2^1A_1$ | R | 5.72 | 6.37 | 6.97 | 7.26 | 7.53 | 8.06 | 8.57 | 7.41 |
| | $2^1B_2$ | R | 6.04 | 6.69 | 7.32 | 7.62 | 7.91 | 8.50 | 9.09 | 7.49 |
| | $3^1A_1$ | R | 6.53 | 7.22 | 7.86 | 8.16 | 8.46 | 9.00 | 9.55 | 7.80 |
| | $3^1B_2$ | R | 6.28 | 6.90 | 7.47 | 7.74 | 7.99 | 8.41 | 8.86 | 8.09 |
| | $1^1B_1$ | R | 6.58 | 7.29 | 7.97 | 8.29 | 8.60 | 9.18 | 9.74 | 8.17 |
| ethylene | $1^1B_{3u}$ | R | 6.46 | 6.85 | 7.23 | 7.40 | 7.58 | 7.91 | 8.04 | 7.11 |
| | $1^1B_{1u}$ | V | 5.65 | 5.73 | 5.81 | 5.84 | 5.88 | 5.95 | 5.98 | 7.65 |
| | $1^1B_{1g}$ | R | 7.01 | 7.47 | 7.91 | 8.12 | 8.32 | 8.70 | 8.91 | 7.80 |
| | $1^1B_{2g}$ | R | 6.95 | 7.43 | 7.88 | 8.09 | 8.31 | 8.72 | 8.95 | 7.90 |
| | $2^1A_g$ | R | 7.33 | 7.73 | 8.12 | 8.31 | 8.49 | 8.86 | 9.14 | 8.28 |
| | $2^1B_{3u}$ | R | 7.79 | 8.28 | 8.71 | 8.89 | 9.07 | 9.39 | 9.59 | 8.62 |
| | $3^1B_{3u}$ | R | 7.99 | 8.42 | 8.88 | 9.12 | 9.36 | 9.82 | 10.23 | 8.90 |
| | $4^1B_{3u}$ | R | 8.26 | 8.75 | 9.25 | 9.50 | 9.74 | 10.23 | 10.78 | 9.08 |
| | $3^1B_{1g}$ | R | 8.55 | 9.01 | 9.45 | 9.67 | 9.88 | 10.28 | 10.70 | 9.20 |
| | $2^1B_{1u}$ | R | 8.12 | 8.60 | 9.08 | 9.31 | 9.55 | 10.01 | 10.54 | 9.33 |
| | $5^1B_{3u}$ | R | 9.30 | 9.74 | 10.15 | 10.34 | 10.53 | 10.91 | 11.37 | 9.51 |
| isobutene | $^1B_1$ | R | 5.37 | 5.74 | 6.08 | 6.24 | 6.40 | 6.70 | 6.84 | 6.17 |

| | | | | | | | | | |
|---|---|---|---|---|---|---|---|---|---|
| | $^1A_1$ | R | 5.23 | 5.34 | 5.43 | 5.47 | 5.51 | 5.58 | 5.65 | 6.7 |
| trans-butadiene | $1^1B_u$ | V | 3.93 | 3.98 | 4.04 | 4.06 | 4.09 | 4.14 | 4.18 | 5.91 |
| | $1^1B_g$ | R | 5.55 | 5.93 | 6.28 | 6.45 | 6.62 | 6.93 | 7.03 | 6.22 |
| | $2^1A_u$ | R | 5.84 | 6.24 | 6.63 | 6.78 | 7.00 | 7.34 | 7.49 | 6.66 |
| | $2^1B_u$ | R | 6.35 | 6.71 | 7.06 | 7.23 | 7.40 | 7.74 | 7.96 | 7.07 |
| | $2^1B_g$ | R | 6.53 | 6.99 | 7.42 | 7.63 | 7.84 | 8.21 | 8.36 | 7.36 |
| | $3^1A_g$ | R | 6.86 | 7.23 | 7.60 | 7.78 | 7.96 | 8.32 | 9.29 | 7.62 |
| | $3^1B_u$ | R | 7.52 | 7.96 | 8.38 | 8.59 | 8.80 | 9.21 | 9.62 | 8.00 |
| pyridine | $^1B_1$ | V | 3.99 | 4.08 | 4.18 | 4.22 | 4.27 | 4.35 | 4.62 | 4.59 |
| | $^1B_2$ | V | 4.82 | 4.83 | 4.84 | 4.85 | 4.86 | 4.87 | 4.85 | 4.99 |
| | $^1A_2$ | V | 4.36 | 4.48 | 4.59 | 4.65 | 4.71 | 4.81 | 5.11 | 5.43 |
| | $^1A_1$ | V | 5.19 | 5.23 | 5.26 | 5.28 | 5.30 | 5.33 | 5.34 | 6.38 |
| pyrazine | $^1B_{3u}$ | V | 3.19 | 3.25 | 3.31 | 3.33 | 3.36 | 3.41 | 3.66 | 3.83 |
| | $^1B_{2u}$ | V | 4.46 | 4.45 | 4.44 | 4.44 | 4.43 | 4.42 | 4.38 | 4.81 |
| | $^1B_{2g}$ | V | 4.67 | 4.76 | 4.84 | 4.89 | 4.93 | 5.01 | 5.22 | 5.46 |
| | $^1B_{1g}$ | V | 5.45 | 5.59 | 5.72 | 5.79 | 5.85 | 5.98 | 6.25 | 6.10 |
| | $^1B_{1u}$ | V | 5.71 | 5.75 | 5.79 | 5.81 | 5.83 | 5.87 | 5.86 | 6.51 |
| pyridazine | $^1B_1$ | V | 2.69 | 2.80 | 2.89 | 2.94 | 2.99 | 3.08 | 3.37 | 3.60 |
| | $^1A_1$ | V | 4.87 | 4.89 | 4.90 | 4.91 | 4.91 | 4.93 | 4.91 | 5.00 |
| | $^1A_2$ | V | 4.68 | 4.76 | 4.84 | 4.88 | 4.92 | 4.98 | 5.22 | 5.30 |
| | $^1B_1$ | V | 5.27 | 5.38 | 5.48 | 5.53 | 5.58 | 5.67 | 5.94 | 6.00 |
| | $^1B_2$ | V | 5.46 | 5.50 | 5.54 | 5.56 | 5.58 | 5.61 | 5.63 | 6.50 |
| pyrimidine | $^1B_1$ | V | 3.48 | 3.58 | 3.67 | 3.72 | 3.76 | 3.85 | 4.13 | 3.85 |
| | $^1A_2$ | V | 3.84 | 3.96 | 4.07 | 4.13 | 4.18 | 4.29 | 4.60 | 4.62 |
| | $^1B_2$ | V | 5.00 | 5.02 | 5.04 | 5.04 | 5.05 | 5.07 | 5.07 | 5.12 |
| | $^1A_2$ | V | 4.73 | 4.82 | 4.91 | 4.95 | 5.00 | 5.08 | 5.34 | 5.52 |

|  | $^1B_1$ | V | 5.08 | 5.20 | 5.31 | 5.36 | 5.42 | 5.52 | 5.80 | 5.90 |
|---|---|---|---|---|---|---|---|---|---|---|
|  | $^1A_1$ | V | 5.36 | 5.40 | 5.44 | 5.45 | 5.47 | 5.51 | 5.53 | 6.70 |
| s-tetrazine | $^1B_{3u}$ | V | 1.42 | 1.49 | 1.55 | 1.58 | 1.61 | 1.67 | 1.94 | 2.25 |
|  | $^1A_u$ | V | 2.64 | 2.76 | 2.87 | 2.93 | 2.98 | 3.09 | 3.43 | 3.40 |
|  | $^1A_u$ | V | 4.25 | 4.29 | 4.33 | 4.34 | 4.36 | 4.39 | 4.59 | 5.00 |
|  | $^1B_{3u}$ | V | 5.47 | 5.56 | 5.65 | 5.69 | 5.73 | 5.81 | 6.08 | 6.34 |
| Mean absolute error (total) |  |  | 1.02 | 0.69 | 0.41 | 0.40 | 0.50 | 0.73 | 0.97 |  |
| Mean absolute error (valence) |  |  | 0.81 | 0.73 | 0.65 | 0.62 | 0.58 | 0.52 | 0.44 |  |
| Mean absolute error (Rydberg) |  |  | 1.18 | 0.65 | 0.23 | 0.23 | 0.43 | 0.89 | 1.38 |  |

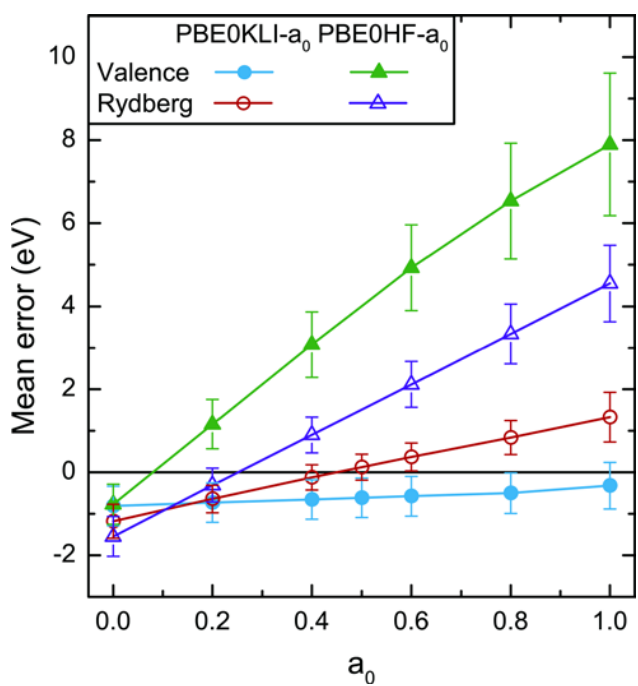

**Figure 3.** The mean error of the excitation energies computed with PBE0HF-$a_0$ and PBE0KLI-$a_0$ orbital energy gaps with respect to the experimental values. The excitation energies computed with PBE0HF-$a_0$ are available in Table S13. The error bars mean standard deviations.

Figure 3 shows the mean error of the excitation energies approximated from the orbital energy gaps with respect to the experimental values in Table 2. It has been known that pure PBE calculations ($a_0 = 0$) underestimate band gap energies, as also shown in Figure 3. In the case of PBE0HF, as $a_0$ increases, the virtual orbital energies increase, whereas the occupied ones decrease (see Figure 1b), resulting in the considerable increase of their energy gaps for both valence and Rydberg excitation energies. In contrast, for the same variation of $a_0$, PBE0KLI gives almost constant valence excitation energies and slight increase of the Rydberg excitation energies, since both occupied and valence orbital energies decrease simultaneously. These trends in PBE0KLI can be further clarified as follows. When the exact exchange potential is added to conventional local or semi-local potentials such as PBE, the energies of occupied and valence virtual orbitals decrease with similar extents, since they are located in the same molecular region. On the other hand, Rydberg orbitals spread mostly outside the molecular region and thus their energies decrease less sensitively compared to those of occupied orbitals.[72] As a result, the energy gaps for the Rydberg excitations are more sensitive to $a_0$ than those for the valence excitation energies. In addition, the standard deviations of PBE0KLI are smaller than those of PBE0HF, which means that the orbital energy gaps of PBE0KLI are distributed closely to the experimental results.

In Table 2 and Figure 3, PBE0KLI-0.5 shows the best result whose mean absolute error (MAE) is 0.40 eV. Therefore, we performed TDDFT calculations with the PBE0KLI-0.5 orbitals. However, the TDDFT kernel of PBE0KLI may not be practical for applications due to high computational costs, and hence we simply used a local density approximation (LDA) kernel[77] on top of the PBE0KLI-0.5 orbitals. Figure 4 and Table S14 show the results. The correction from the LDA kernel improved the valence excitation energies, leading to the significant decrease of MAE from 0.62 eV to 0.35 eV, while it is almost ineffective for the Rydberg excitations (Figure 5 and Table S14). The latter is because the overlap between occupied and virtual orbitals for the Rydberg excitations is so small that the correction from the LDA kernel is negligible due to its locality (red squares in Figure 4).[78] On the other hand, occupied and virtual orbitals corresponding to the valence excitations significantly overlap with one another, and thus the LDA kernel improves the excitation energies (blue circles in Figure 4).

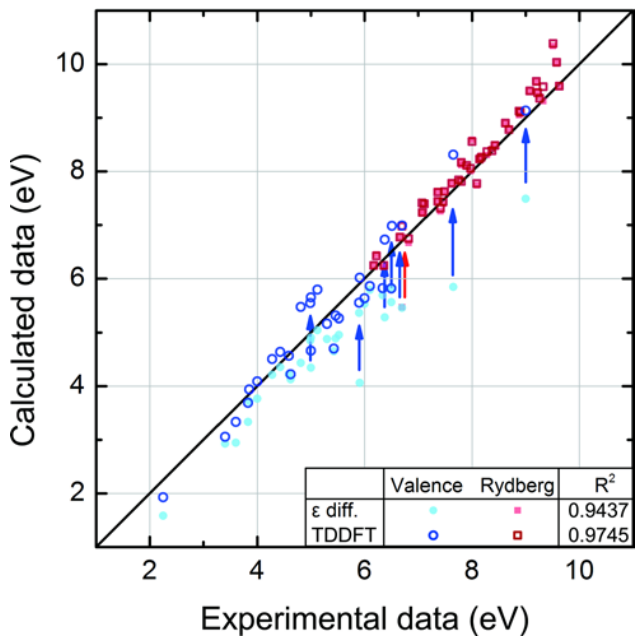

**Figure 4.** Comparison between calculated and experimental excitation energies for the Caricato set. PBE0KLI-0.5 has been used. The filled and empty symbols indicate the values from orbital energy gaps and TDDFT calculations with the LDA kernel, respectively. The blue and red arrows denote the noticeable improvement of excitation energies by the LDA kernel.

**Table 3.** The excitation energies of Caricato set calculated using the PBE0KLI-0.5 functional. This table only includes the excitations corrected significantly by the LDA kernel (> 1.0 eV). The third column indicates excitation types; V and R stand for vertical and Rydberg excitations, respectively. Experimental values were obtained from Ref. 63. Unit: eV

| Molecule | State | Type | PBE0KLI-0.5 | | the portion of the major configuration in TDDFT ($|c|^2$) | Expt. |
| --- | --- | --- | --- | --- | --- | --- |
| | | | Orbital energy gaps | TDDFT w/ LDA kernel | | |
| formaldehyde | $1^1B_1$ | V | 7.49 | 9.14 | 0.92 | 9.00 |
| ethylene | $1^1B_{1u}$ | V | 5.84 | 8.31 | 0.73 | 7.65 |
| isobutene | $^1A_1$ | R | 5.47 | 6.98 | 0.49 | 6.70 |
| trans-butadiene | $1^1B_u$ | V | 4.06 | 6.02 | 0.85 | 5.91 |
| pyridine | $^1A_1$ | V | 5.28 | 6.73 | 1.00 | 6.38 |
| pyrazine | $^1B_{2u}$ | V | 4.44 | 5.47 | 0.80 | 4.81 |
| | $^1B_{1u}$ | V | 5.81 | 6.98 | 0.65 | 6.51 |
| pyrimidine | $^1A_1$ | V | 5.45 | 6.99 | 0.66 | 6.70 |

In many cases including the Rydberg excitations, single orbital transitions from PBE0KLI-0.5 were sufficient to produce accurate excitation energies as is evident from the high accuracy of the orbital energy gaps. However, the LDA correction becomes essential as multi-configurational effects are indispensable. This argument can be supported by analyzing the portion of the major configuration of TDDFT results ($|c|^2$). The average value of $|c|^2$ for all the excitations in the Caricato set is 0.93 (Table S14). In the case of that the orbital energy gaps well approximate the excitation energies, $|c|^2$ values are almost equal to 1. For some excitation energies, the LDA correction is larger than 1.0 eV as marked by the arrows in Figure 4. Their average $|c|^2$ value is 0.71 (Table 3), meaning that they require more than a single configuration to have accurate excitation energies. Such multi-configuration effects are particularly important for excitations involving degenerate orbitals.

Surprisingly, PBE0KLI-0.5 even with the LDA kernel outperforms all the conventional pure, hybrid, and long-range corrected functionals[63,72,74,79] considered here except for MN15[80] as indicated by the arrows in Figure 5. Such an unexpected high accuracy is not a fortunate coincidence but the consequence of physically meaningful virtual orbitals as discussed in Figure 1. Due to the similar reason, SAOP also with the LDA kernel gives relatively small MAE (0.34 eV). It is designed to mimic the exact KS xc potential such that it satisfies the correct asymptotic behavior and the Koopman's theorem.[72,81,82]

Unlike conventional global hybrid methods, PBE0KLI-0.5 gives very accurate Rydberg excitation energies because of the following two reasons. First, the high EXX ratio improves the correct asymptotic behavior of KS potential.[79] For instance, Isegawa and co-workers reported that the EXX ratio in hybrid functionals should be similar to or larger than 50% for accurate Rydberg excitation energies through substantial mitigation of the self-interaction error. In fact, M06-2X[17] and MN15[80] with EXX of 54% and 44%, respectively, showed the best results for the Rydberg excitation energies among the others in Figure 5. Second, virtual orbitals from PBE0KLI-0.5 form bound states due to their $N$-electron characters and the correct asymptotic behavior of KS potential. Therefore, they closely resemble Rydberg states.

Since we used two different types of basis set for excited state calculations, we need to investigate the basis set size effect on the MAE values. For the same PBE functional, MAE from the Gaussian basis set is larger than that of the Lagrange-sinc basis set that is indicated by the dashed arrow in Figure 5. In particular, the Lagrange-sinc basis set has smaller errors for the Rydberg excitations than the Gaussian basis set, whereas it has larger errors for the valence excitations. The former seems natural because the Lagrange-sinc basis set adopts a numerical grid so that it can well describe diffuse orbitals. The latter may be due to the error cancellation effect of the Gaussian basis set and the use of pseudopotentials in the Lagrange-sinc basis set. However, such effects do not change the above distinctive features of KLI hybrid.

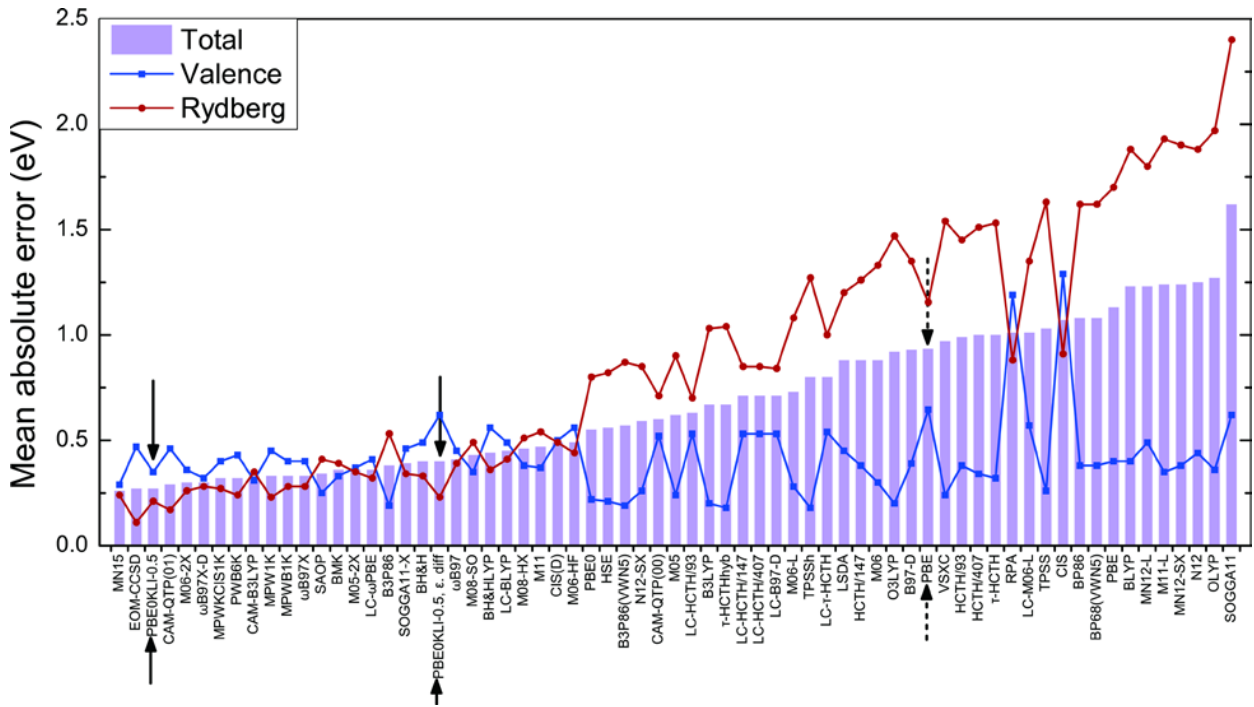

**Figure 5.** The MAE of the 69 excitation energies in the Caricato set[63] for various computational methods. The results from PBE0KLI are indicated by the two black arrows. The dashed arrow denotes the MAE value from PBE with the Lagrange-sinc basis set. Most MAE values were obtained from the original Caricato's paper (Ref. 63) and the benchmark data of Isegawa and co-workers.[79] The MAE values of CAM-QTP(00)[73] and CAM-QTP(01)[74] are in Ref. 74, that of SAOP is in Ref. 72, and that of MN15 is in Ref. 80.

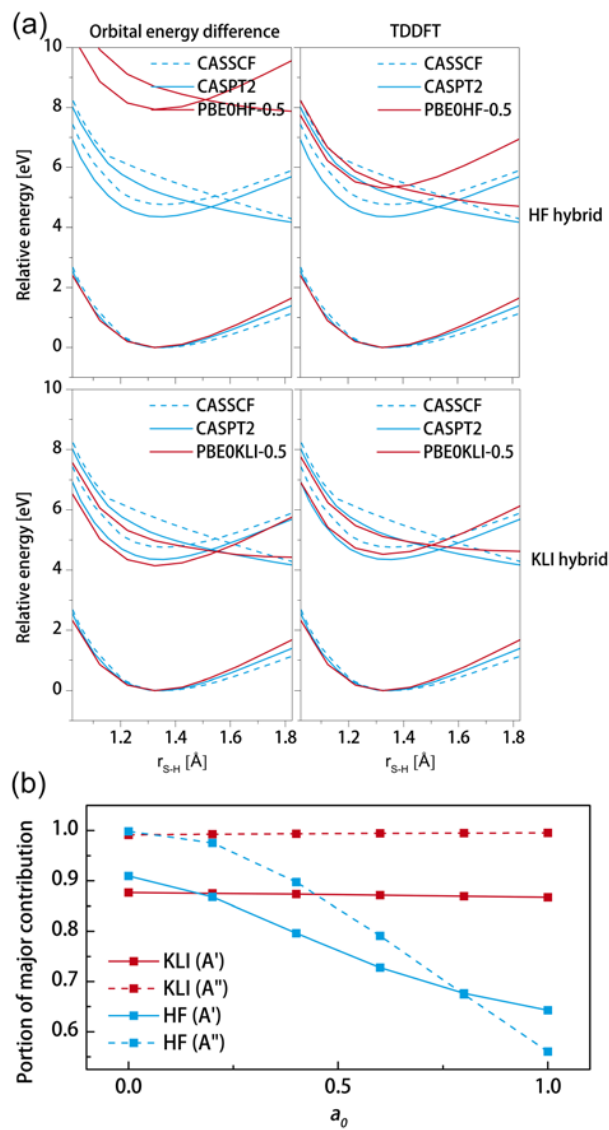

**Figure 6.** (a) The potential energy curves computed with CASSCF, CASPT2, PBE0HF-0.5, and PBE0KLI-0.5. The left panels correspond to the energy gaps between occupied and virtual orbitals and the right ones are the corresponding TDDFT results with the LDA kernel. (b) The change in the portion of the major configuration (the square of the normalized coefficient) in TDDFT at $r_{S-H}=1.32$Å with increasing mixing ratio.

**Application to thiophenol photodissociation.** In this section, we computed the potential energy curves (PECs) of thiophenol near the ground-state geometry as a function of the distance between S and H using TDDFT and compared the result with those of CASSCF and CASPT2. This system was chosen because it has been widely studied experimentally and theoretically to understand non-adiabatic chemical dynamics through conical intersections.[83–87] The left panels in Figure 6a show the energy gaps between occupied and virtual orbitals corresponding to each excitation. The energy gaps of PBE0HF-0.5 are too large to approximate the excitation energies of CASPT2 (upper panel). The subsequent TDDFT correction significantly improves the PECs, but the position of the conical intersection ($r_{S-H}$ = 1.38 Å) is noticeably different from that of CASPT2 ($r_{S-H}$ = 1.53 Å) as shown in the right-upper panel in Figure 6a. The variation of $a_0$ does not improve the result (Figure S2). In contrast, the orbital energy gaps of PBE0KLI-0.5 are already close to the PECs of CASPT2 (left-lower panel), and the TDDFT correction with the LDA kernel further improves the result, and especially the position of the conical intersection ($r_{S-H}$ = 1.49 Å) is very close to that of CASPT2 (right-lower panel).

It has been discussed that KS orbital energy gaps approximate optical excitation energies reasonably, while HF orbital energy gaps correspond to fundamental energy gaps, i.e., half the chemical hardness.[13] Figure 6b shows the change in the portion of the major configuration in TDDFT as the mixing ratio increases. For both A' and A" excitations, that of the KLI hybrid remains almost constant, manifesting the characteristic of single orbital transitions, whereas that of the HF hybrid continuously decreases. This is because virtual orbitals of the latter with a high mixing ratio are delocalized (Figure S3), so that more configurations are required to correct excitation energies.

**CONCLUSIONS**

The present work offers a new perspective of hybrid functionals: local vs non-local. Formally, the local version is more desirable because it satisfies the locality of Kohn-Sham theory. In practice, it describes ground states as accurately as the standard one; both show similar results for atomization energies and barrier heights. However, they give different results for excited states. For example, in the benchmark calculations for the 30 valence and 39 Rydberg excitations in the Caricato set, the orbital energy gaps from PBE0 with a local exact exchange potential well approximate excitation energies with a small MAE (0.40 eV) which is already better than various TDDFT results (Figure 5). In particular, it shows higher accuracy for Rydberg excitations. TDDFT even with a LDA kernel significantly improves accuracy for valence excitation energies by incorporating multi-configurational effects. As a result, its MAE reduces down to 0.27 eV which is smaller than those of all the other pure and hybrid functionals except for MN15. In most cases, single orbital transitions are sufficient to give accurate excitation energies as is evident from the portion of major configurations in TDDFT close to 1. As an application, the local version of PBE0 shows comparable accuracy with CASPT2 for the potential energy curves of thiophenol photodissociation.

It is surprising to note that the local hybrid method even with a LDA kernel outperforms conventional functionals for excitation energy calculations. It shows high accuracy for Rydberg excitations for which conventional ones often fail. However, it is doubtable whether it also works well for charge-transfer excitations due to the locality of the LDA kernel. It is well known that long-range charge-transfer excitations can be described as follows[88]:

$$\text{(Excitation energy)} \cong \text{(Ionization Potential)} - \text{(Electron Affinity)} - \frac{1}{R}, \qquad (24)$$

where $R$ is the distance between donor and acceptor regions. That cannot be approximated by simply orbital energy gaps. For the general applicability of the local hybrid method, a long-range corrected hybrid functional[25] should be employed.

It should be emphasized that the hybrid method with a local exact exchange potential is not a new type of functionals, because it adopts conventional hybrid functional forms without any parameter re-adjustment. Instead, it utilizes a *local* potential to incorporate the effect of exact exchange within the KS picture, whereas conventional hybrids use the *non-local* Fock operator justified by the GKS theory. We noted that such a locality plays an important role for accurate excited states. Apparently, this perspective offers a new way to improve the accuracy of hybrid functionals for excited states.

AUTHOR INFORMATION

**Corresponding Author**

*E-mail: wooyoun@kaist.ac.kr

**Author Contributions**

†These authors contributed equally.

**Notes**

The authors declare no competing financial interests.

ACKNOWLEDGMENT

This work was supported by Basic Science Research Programs (NRF-2015R1A1A1A05001480) and EDISON Program (NRF-2012M3C1A6035359) funded by the Korea government [MSIP]. The authors would like to acknowledge the support from KISTI supercomputing center (No.

KSC-2015-C2-043). W.Y.K. is grateful for the financial support from EWon associate professorship. We are also grateful to Profs. Evert J. Barerends, Kieron Burke, Kwang S. Kim, and Yoon Sup Lee for their valuable comments.

Supporting Information for

# Importance of local exact exchange potential in hybrid functionals for accurate excited states

Jaewook Kim,† Kwangwoo Hong,†
Sang-Yeon Hwang, Seongok Ryu, Sunghwan Choi, and Woo Youn Kim*

Department of Chemistry, KAIST, 291 Daehak-ro, Yuseong-gu, Daejeon 34141, Republic of Korea

†Jaewook Kim and Kwangwoo Hong contributed equally to this work.

*E-mail: wooyoun@kaist.ac.kr

# 1. Ground state geometry of thiophenol

**Table S1.** The Cartesian coordinate of ground-state thiophenol in Angstrom.

|   | x | y | z |
|---|---|---|---|
| C | 0.000 | 0.000 | 0.000 |
| C | 0.000 | 0.000 | 1.393 |
| C | 1.212 | 0.000 | 2.083 |
| C | 2.412 | 0.000 | 1.382 |
| C | 2.415 | 0.000 | -0.010 |
| C | 1.205 | 0.000 | -0.695 |
| H | -0.925 | 0.000 | -0.543 |
| H | 1.220 | 0.000 | 3.156 |
| H | 3.339 | 0.000 | 1.922 |
| H | 3.342 | 0.000 | -0.549 |
| H | 1.194 | 0.000 | -1.768 |
| S | -1.516 | 0.000 | 2.366 |
| H | -2.372 | 0.000 | 1.355 |

## 2. Geometries used for Caricato set calculations

**Table S2.** The Cartesian coordinate of formaldehyde in Angstrom.

|   | x | y | z |
|---|---|---|---|
| H | 0.000000 | 0.934473 | -0.588078 |
| H | 0.000000 | -0.934473 | -0.588078 |
| C | 0.000000 | 0.000000 | 0.000000 |
| O | 0.000000 | 0.000000 | 1.221104 |

**Table S3.** The Cartesian coordinate of acetaldehyde in Angstrom.

|   | x | y | z |
|---|---|---|---|
| O | 1.212008 | 0.374458 | 0.000000 |
| C | 0.000000 | 0.462805 | 0.000000 |
| H | -0.486928 | 1.460337 | 0.000000 |
| C | -0.941279 | -0.711815 | 0.000000 |
| H | -0.384684 | -1.649523 | 0.000000 |
| H | -1.588387 | -0.656210 | 0.881703 |
| H | -1.588387 | -0.656210 | -0.881703 |

**Table S4.** The Cartesian coordinate of acetone in Angstrom.

|   | x | y | z |
|---|---|---|---|
| H | 0.000000 | 2.136732 | -0.112445 |
| H | 0.000000 | -2.136732 | -0.112445 |
| H | -0.881334 | 1.333733 | -1.443842 |
| H | 0.881334 | -1.333733 | -1.443842 |
| H | -0.881334 | -1.333733 | -1.443842 |
| H | 0.881334 | 1.333733 | -1.443842 |
| C | 0.000000 | 0.000000 | 0.000000 |
| C | 0.000000 | 1.287253 | -0.795902 |
| C | 0.000000 | -1.287253 | -0.795902 |
| O | 0.000000 | 0.000000 | 1.227600 |

**Table S5.** The Cartesian coordinate of ethylene in Angstrom.

|   | x | y | z |
|---|---|---|---|
| H | 0.000000 | 0.923274 | 1.238289 |
| H | 0.000000 | -0.923274 | 1.238289 |
| H | 0.000000 | 0.923274 | -1.238289 |
| H | 0.000000 | -0.923274 | -1.238289 |
| C | 0.000000 | 0.000000 | 0.668188 |
| C | 0.000000 | 0.000000 | -0.668188 |

**Table S6.** The Cartesian coordinate of isobutene in Angstrom.

|   | x | y | z |
|---|---|---|---|
| C | 0.000000 | 0.000000 | 1.463400 |
| C | 0.000000 | 0.000000 | 0.119614 |
| H | 0.000000 | 0.928447 | 2.027281 |
| H | 0.000000 | -0.928447 | 2.027281 |
| C | 0.000000 | 1.276163 | -0.680400 |
| H | 0.000000 | 2.156688 | -0.032951 |
| H | 0.881626 | 1.321381 | -1.330486 |
| H | -0.881626 | 1.321381 | -1.330486 |
| C | 0.000000 | -1.276163 | -0.680400 |
| H | 0.000000 | -2.156688 | -0.032951 |
| H | -0.881626 | -1.321381 | -1.330486 |
| H | 0.881626 | -1.321381 | -1.330486 |

**Table S7.** The Cartesian coordinate of trans-butadiene in Angstrom.

|   | x | y | z |
|---|---|---|---|
| H | 1.080977 | -2.558832 | 0.000000 |
| H | -1.080977 | 2.558832 | 0.000000 |
| H | 2.103773 | -1.017723 | 0.000000 |
| H | -2.103773 | 1.017723 | 0.000000 |
| H | -0.973565 | -1.219040 | 0.000000 |
| H | 0.973565 | 1.219040 | 0.000000 |
| C | 0.000000 | 0.728881 | 0.000000 |
| C | 0.000000 | -0.728881 | 0.000000 |
| C | 1.117962 | -1.474815 | 0.000000 |
| C | -1.117962 | 1.474815 | 0.000000 |

**Table S8.** The Cartesian coordinate of pyridine in Angstrom.

|   | x | y | z |
|---|---|---|---|
| H | 0.000000 | 2.061947 | 1.308539 |
| H | 0.000000 | -2.061947 | 1.308539 |
| H | 0.000000 | 2.156804 | -1.184054 |
| H | 0.000000 | -2.156804 | -1.184054 |
| H | 0.000000 | 0.000000 | -2.475074 |
| C | 0.000000 | 1.145417 | 0.721005 |
| C | 0.000000 | -1.145417 | 0.721005 |
| C | 0.000000 | 1.197637 | -0.673735 |
| C | 0.000000 | -1.197637 | -0.673735 |
| C | 0.000000 | 0.000000 | -1.387901 |
| N | 0.000000 | 0.000000 | 1.426610 |

**Table S9.** The Cartesian coordinate of pyrazine in Angstrom.

|   | x | y | z |
|---|---|---|---|
| H | 0.000000 | 2.068464 | 1.258236 |
| H | 0.000000 | -2.068464 | 1.258236 |
| H | 0.000000 | 2.068464 | -1.258236 |
| H | 0.000000 | -2.068464 | -1.258236 |
| C | 0.000000 | 1.135920 | 0.697884 |
| C | 0.000000 | -1.135920 | 0.697884 |
| C | 0.000000 | 1.135920 | -0.697884 |
| C | 0.000000 | -1.135920 | -0.697884 |
| N | 0.000000 | 0.000000 | 1.417402 |
| N | 0.000000 | 0.000000 | -1.417402 |

**Table S10.** The Cartesian coordinate of pyridazine in Angstrom.

|   | x | y | z |
|---|---|---|---|
| H | 0.000000 | 2.409486 | -0.149325 |
| H | 0.000000 | -2.409486 | -0.149325 |
| H | 0.000000 | 1.271234 | 2.102647 |
| H | 0.000000 | -1.271234 | 2.102647 |
| C | 0.000000 | 1.325698 | -0.063084 |
| C | 0.000000 | -1.325698 | -0.063084 |
| C | 0.000000 | 0.693095 | 1.182948 |
| C | 0.000000 | -0.693095 | 1.182948 |
| N | 0.000000 | 0.674211 | -1.238929 |
| N | 0.000000 | -0.674211 | -1.238929 |

**Table S11.** The Cartesian coordinate of pyrimidine in Angstrom.

|   | x | y | z |
|---|---|---|---|
| H | 0.000000 | 2.156588 | 1.120200 |
| H | 0.000000 | -2.156588 | 1.120200 |
| H | 0.000000 | 0.000000 | -2.400385 |
| H | 0.000000 | 0.000000 | 2.440403 |
| C | 0.000000 | 1.186684 | 0.626213 |
| C | 0.000000 | -1.186684 | 0.626213 |
| C | 0.000000 | 0.000000 | -1.312625 |
| C | 0.000000 | 0.000000 | 1.354949 |
| N | 0.000000 | 1.203523 | -0.717781 |
| N | 0.000000 | -1.203523 | -0.717781 |

**Table S12.** The Cartesian coordinate of s-tetrazine in Angstrom.

|   | x | y | z |
|---|---|---|---|
| H | 0.000000 | 0.000000 | -2.354794 |
| H | 0.000000 | 0.000000 | 2.354794 |
| C | 0.000000 | 0.000000 | 1.269044 |
| C | 0.000000 | 0.000000 | -1.269044 |
| N | 0.000000 | 1.204572 | 0.670429 |
| N | 0.000000 | -1.204572 | 0.670429 |
| N | 0.000000 | 1.204572 | -0.670429 |
| N | 0.000000 | -1.204572 | -0.670429 |

## 3. Atomization energy

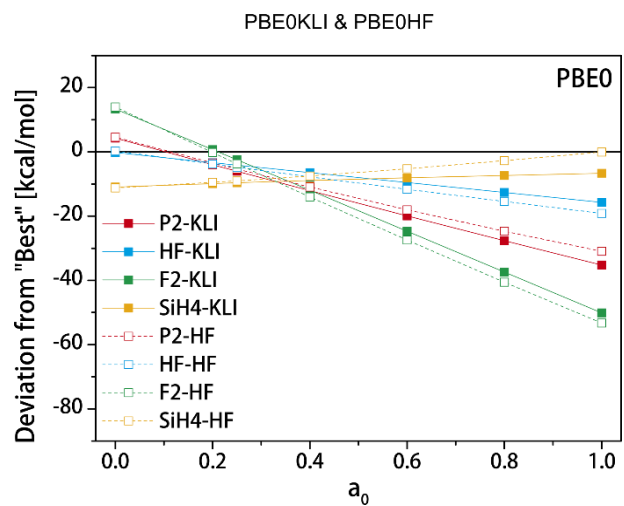

**Figure S1.** Atomization energies of $P_2$, HF, $F_2$, and $SiH_4$ as a function of the exact exchange ratio ($a_0$) with respect to the ab initio calculations [R. Haunschild, W. Klopper, *J. Chem. Phys.* **2012**, *136*, 164102] which are denoted as "Best".

## 4. Caricato set - Orbital energy gaps from PBE0HF-$a_0$

**Table S13.** The orbital energy gaps from PBE0HF-$a_0$ with various $a_0$ as an approximation to each excitation energy for the Caricato set. The third column indicates the type of excitations: V and R for valence and Rydberg excitations, respectively. The experimental values were obtained from [M. Caricato, G. W. Trucks, M. J. Frisch, K. B. Wiberg, *J. Chem. Theory Comput.* **2010**, *6* (2), 370–383]. Unit: eV

| Molecule | State | Type | PBE0HF-$a_0$ (Orbital energy difference) | | | | | | Expt. |
|---|---|---|---|---|---|---|---|---|---|
| | | | 0.0 | 0.2 | 0.4 | 0.6 | 0.8 | 1.0 | |
| formaldehyde | $1^1A_2$ | V | 3.46 | 3.59 | 3.71 | 3.83 | 3.95 | 4.28 | 4.00 |
| | $1^1B_2$ | R | 5.83 | 6.41 | 6.96 | 7.48 | 7.98 | 8.49 | 7.08 |
| | $2^1B_2$ | R | 6.67 | 7.23 | 7.76 | 8.24 | 8.69 | 9.17 | 7.97 |
| | $2^1A_1$ | R | 6.61 | 7.28 | 7.92 | 8.52 | 9.10 | 9.75 | 8.14 |
| | $2^1A_2$ | R | 6.92 | 7.53 | 8.13 | 8.70 | 9.27 | 10.00 | 8.37 |
| | $3^1B_2$ | R | 7.46 | 8.13 | 8.75 | 9.34 | 9.88 | 10.55 | 8.88 |
| | $1^1B_1$ | V | 8.07 | 8.24 | 8.40 | 8.56 | 8.71 | 9.00 | 9.00 |
| | $3^1A_2$ | R | 7.73 | 8.44 | 9.14 | 9.84 | 10.54 | 11.52 | 9.22 |
| | $4^1B_2$ | R | 7.66 | 8.36 | 9.05 | 9.73 | 10.39 | 11.27 | 9.26 |
| | $4^1A_1$ | R | 8.23 | 8.94 | 9.66 | 10.38 | 11.10 | 12.20 | 9.58 |
| | $5^1B_2$ | R | 7.71 | 8.46 | 9.21 | 9.97 | 10.73 | 11.86 | 9.63 |
| acetaldehyde | $1^1A''$ | V | 3.84 | 3.99 | 4.14 | 4.28 | 4.42 | 4.75 | 4.28 |
| | $2^1A'$ | R | 5.38 | 5.92 | 6.42 | 6.90 | 7.34 | 7.78 | 6.82 |
| | $3^1A'$ | R | 5.94 | 6.55 | 7.12 | 7.66 | 8.16 | 8.66 | 7.46 |
| | $4^1A'$ | R | 6.33 | 6.96 | 7.55 | 8.09 | 8.60 | 9.10 | 7.75 |
| | $6^1A'$ | R | 6.90 | 7.56 | 8.19 | 8.79 | 9.35 | 9.98 | 8.43 |
| | $7^1A'$ | R | 7.07 | 7.78 | 8.47 | 9.13 | 9.76 | 10.51 | 8.69 |
| acetone | $1^1A_2$ | V | 3.93 | 4.11 | 4.27 | 4.44 | 4.59 | 4.91 | 4.43 |
| | $1^1B_2$ | R | 4.96 | 5.49 | 5.98 | 6.44 | 6.87 | 7.27 | 6.36 |
| | $2^1A_2$ | R | 5.93 | 6.56 | 7.16 | 7.72 | 8.24 | 8.77 | 7.36 |
| | $2^1A_1$ | R | 5.72 | 6.37 | 6.97 | 7.53 | 8.06 | 8.57 | 7.41 |
| | $2^1B_2$ | R | 6.04 | 6.69 | 7.32 | 7.91 | 8.50 | 9.09 | 7.49 |
| | $3^1A_1$ | R | 6.53 | 7.22 | 7.86 | 8.46 | 9.00 | 9.55 | 7.80 |
| | $3^1B_2$ | R | 6.28 | 6.90 | 7.47 | 7.99 | 8.41 | 8.86 | 8.09 |
| | $1^1B_1$ | R | 6.58 | 7.29 | 7.97 | 8.60 | 9.18 | 9.74 | 8.17 |
| ethylene | $1^1B_{3u}$ | R | 6.46 | 6.85 | 7.23 | 7.58 | 7.91 | 8.04 | 7.11 |
| | $1^1B_{1u}$ | V | 5.65 | 5.73 | 5.81 | 5.88 | 5.95 | 5.98 | 7.65 |
| | $1^1B_{1g}$ | R | 7.01 | 7.47 | 7.91 | 8.32 | 8.70 | 8.91 | 7.80 |
| | $1^1B_{2g}$ | R | 6.95 | 7.43 | 7.88 | 8.31 | 8.72 | 8.95 | 7.90 |
| | $2^1A_g$ | R | 7.33 | 7.73 | 8.12 | 8.49 | 8.86 | 9.14 | 8.28 |
| | $2^1B_{3u}$ | R | 7.79 | 8.28 | 8.71 | 9.07 | 9.39 | 9.59 | 8.62 |
| | $3^1B_{3u}$ | R | 7.99 | 8.42 | 8.88 | 9.36 | 9.82 | 10.23 | 8.90 |
| | $4^1B_{3u}$ | R | 8.26 | 8.75 | 9.25 | 9.74 | 10.23 | 10.78 | 9.08 |
| | $3^1B_{1g}$ | R | 8.55 | 9.01 | 9.45 | 9.88 | 10.28 | 10.70 | 9.20 |
| | $2^1B_{1u}$ | R | 8.12 | 8.60 | 9.08 | 9.55 | 10.01 | 10.54 | 9.33 |
| | $5^1B_{3u}$ | R | 9.30 | 9.74 | 10.15 | 10.53 | 10.91 | 11.37 | 9.51 |
| isobutene | $^1B_1$ | R | 5.37 | 5.74 | 6.08 | 6.40 | 6.70 | 6.84 | 6.17 |
| | $^1A_1$ | R | 5.23 | 5.34 | 5.43 | 5.51 | 5.58 | 5.65 | 6.7 |

| | | | | | | | | |
|---|---|---|---|---|---|---|---|---|
| trans-butadiene | $1^1B_u$ | V | 3.93 | 3.98 | 4.04 | 4.09 | 4.14 | 4.18 | 5.91 |
| | $1^1B_g$ | R | 5.55 | 5.93 | 6.28 | 6.62 | 6.93 | 7.03 | 6.22 |
| | $2^1A_u$ | R | 5.84 | 6.24 | 6.63 | 7.00 | 7.34 | 7.49 | 6.66 |
| | $2^1B_u$ | R | 6.35 | 6.71 | 7.06 | 7.40 | 7.74 | 7.96 | 7.07 |
| | $2^1B_g$ | R | 6.53 | 6.99 | 7.42 | 7.84 | 8.21 | 8.36 | 7.36 |
| | $3^1A_g$ | R | 6.86 | 7.23 | 7.60 | 7.96 | 8.32 | 9.29 | 7.62 |
| | $3^1B_u$ | R | 7.52 | 7.96 | 8.38 | 8.80 | 9.21 | 9.62 | 8.00 |
| pyridine | $^1B_1$ | V | 3.99 | 4.08 | 4.18 | 4.27 | 4.35 | 4.62 | 4.59 |
| | $^1B_2$ | V | 4.82 | 4.83 | 4.84 | 4.86 | 4.87 | 4.85 | 4.99 |
| | $^1A_2$ | V | 4.36 | 4.48 | 4.59 | 4.71 | 4.81 | 5.11 | 5.43 |
| | $^1A_1$ | V | 5.19 | 5.23 | 5.26 | 5.30 | 5.33 | 5.34 | 6.38 |
| pyrazine | $^1B_{3u}$ | V | 3.19 | 3.25 | 3.31 | 3.36 | 3.41 | 3.66 | 3.83 |
| | $^1B_{2u}$ | V | 4.46 | 4.45 | 4.44 | 4.43 | 4.42 | 4.38 | 4.81 |
| | $^1B_{2g}$ | V | 4.67 | 4.76 | 4.84 | 4.93 | 5.01 | 5.22 | 5.46 |
| | $^1B_{1g}$ | V | 5.45 | 5.59 | 5.72 | 5.85 | 5.98 | 6.25 | 6.10 |
| | $^1B_{1u}$ | V | 5.71 | 5.75 | 5.79 | 5.83 | 5.87 | 5.86 | 6.51 |
| pyridazine | $^1B_1$ | V | 2.69 | 2.80 | 2.89 | 2.99 | 3.08 | 3.37 | 3.60 |
| | $^1A_1$ | V | 4.87 | 4.89 | 4.90 | 4.91 | 4.93 | 4.91 | 5.00 |
| | $^1A_2$ | V | 4.68 | 4.76 | 4.84 | 4.92 | 4.98 | 5.22 | 5.30 |
| | $^1B_1$ | V | 5.27 | 5.38 | 5.48 | 5.58 | 5.67 | 5.94 | 6.00 |
| | $^1B_2$ | V | 5.46 | 5.50 | 5.54 | 5.58 | 5.61 | 5.63 | 6.50 |
| pyrimidine | $^1B_1$ | V | 3.48 | 3.58 | 3.67 | 3.76 | 3.85 | 4.13 | 3.85 |
| | $^1A_2$ | V | 3.84 | 3.96 | 4.07 | 4.18 | 4.29 | 4.60 | 4.62 |
| | $^1B_2$ | V | 5.00 | 5.02 | 5.04 | 5.05 | 5.07 | 5.07 | 5.12 |
| | $^1A_2$ | V | 4.73 | 4.82 | 4.91 | 5.00 | 5.08 | 5.34 | 5.52 |
| | $^1B_1$ | V | 5.08 | 5.20 | 5.31 | 5.42 | 5.52 | 5.80 | 5.90 |
| | $^1A_1$ | V | 5.36 | 5.40 | 5.44 | 5.47 | 5.51 | 5.53 | 6.70 |
| s-tetrazine | $^1B_{3u}$ | V | 1.42 | 1.49 | 1.55 | 1.61 | 1.67 | 1.94 | 2.25 |
| | $^1A_u$ | V | 2.64 | 2.76 | 2.87 | 2.98 | 3.09 | 3.43 | 3.40 |
| | $^1A_u$ | V | 4.25 | 4.29 | 4.33 | 4.36 | 4.39 | 4.59 | 5.00 |
| | $^1B_{3u}$ | V | 5.47 | 5.56 | 5.65 | 5.73 | 5.81 | 6.08 | 6.34 |
| Mean absolute error (total) | | | 1.02 | 0.69 | 0.41 | 0.50 | 0.73 | 0.97 | |
| Mean absolute error (valence) | | | 0.81 | 0.73 | 0.65 | 0.58 | 0.52 | 0.44 | |
| Mean absolute error (Rydberg) | | | 1.18 | 0.65 | 0.23 | 0.43 | 0.89 | 1.38 | |

## 4. Caricato set – TDDFT results from PBE0KLI-0.5

**Table S14.** The excitation energies of Caricato set calculated using the PBE0KLI-0.5 functional. The third column indicates excitation types: V and R for vertical and Rydberg excitations, respectively. Experimental values were obtained from [M. Caricato, G. W. Trucks, M. J. Frisch, K. B. Wiberg, *J. Chem. Theory Comput.* **2010**, *6* (2), 370–383]. Unit: eV

| Molecule | State | Type | PBE0KLI-0.5 (Orbital energy gaps) | PBE0KLI-0.5 w/ LDA kernel (TDDFT) | the portion of the major configuration in TDDFT calculation ($|c|^2$) | Expt. |
|---|---|---|---|---|---|---|
| formaldehyde | $1^1A_2$ | V | 3.77 | 4.09 | 1.00 | 4.00 |
| | $1^1B_2$ | R | 7.22 | 7.23 | 1.00 | 7.08 |
| | $2^1B_2$ | R | 8.00 | 8.05 | 0.99 | 7.97 |
| | $2^1A_1$ | R | 8.22 | 8.23 | 0.99 | 8.14 |
| | $2^1A_2$ | R | 8.42 | 8.38 | 1.00 | 8.37 |
| | $3^1B_2$ | R | 9.05 | 9.12 | 0.97 | 8.88 |
| | $1^1B_1$ | V | 7.49 | 9.14 | 0.92 | 9.00 |
| | $3^1A_2$ | R | 9.50 | 9.47 | 1.00 | 9.22 |
| | $4^1B_2$ | R | 9.39 | 9.36 | 1.00 | 9.26 |
| | $4^1A_1$ | R | 10.02 | 10.03 | 0.92 | 9.58 |
| | $5^1B_2$ | R | 9.59 | 9.59 | 0.99 | 9.63 |
| acetaldehyde | $1^1A''$ | V | 4.21 | 3.99 | 1.00 | 4.28 |
| | $2^1A'$ | R | 6.66 | 5.92 | 0.99 | 6.82 |
| | $3^1A'$ | R | 7.39 | 6.55 | 0.99 | 7.46 |
| | $4^1A'$ | R | 7.82 | 6.96 | 0.98 | 7.75 |
| | $6^1A'$ | R | 8.50 | 7.56 | 0.99 | 8.43 |
| | $7^1A'$ | R | 8.80 | 7.78 | 0.99 | 8.69 |
| acetone | $1^1A_2$ | V | 4.35 | 4.64 | 1.00 | 4.43 |
| | $1^1B_2$ | R | 6.21 | 6.24 | 1.00 | 6.36 |
| | $2^1A_2$ | R | 7.44 | 7.44 | 0.99 | 7.36 |
| | $2^1A_1$ | R | 7.26 | 7.31 | 0.97 | 7.41 |
| | $2^1B_2$ | R | 7.62 | 7.62 | 0.99 | 7.49 |
| | $3^1A_1$ | R | 8.16 | 8.17 | 1.00 | 7.80 |
| | $3^1B_2$ | R | 7.74 | 7.77 | 0.99 | 8.09 |
| | $1^1B_1$ | R | 8.29 | 8.26 | 0.97 | 8.17 |
| ethylene | $1^1B_{3u}$ | R | 7.40 | 7.40 | 1.00 | 7.11 |
| | $1^1B_{1u}$ | V | 5.84 | 8.31 | 0.73 | 7.65 |
| | $1^1B_{1g}$ | R | 8.12 | 7.81 | 0.96 | 7.80 |
| | $1^1B_{2g}$ | R | 8.09 | 8.11 | 1.00 | 7.90 |
| | $2^1A_g$ | R | 8.31 | 8.37 | 0.99 | 8.28 |
| | $2^1B_{3u}$ | R | 8.89 | 8.90 | 0.99 | 8.62 |
| | $3^1B_{3u}$ | R | 9.12 | 9.10 | 0.99 | 8.90 |
| | $4^1B_{3u}$ | R | 9.50 | 9.50 | 1.00 | 9.08 |
| | $3^1B_{1g}$ | R | 9.67 | 9.68 | 1.00 | 9.20 |
| | $2^1B_{1u}$ | R | 9.31 | 9.58 | 0.72 | 9.33 |
| | $5^1B_{3u}$ | R | 10.34 | 10.38 | 0.91 | 9.51 |
| isobutene | $^1B_1$ | R | 6.24 | 6.25 | 1.00 | 6.17 |
| | $^1A_1$ | R | 5.47 | 6.98 | 0.49 | 6.70 |
| trans-butadiene | $1^1B_u$ | V | 4.06 | 6.02 | 0.85 | 5.91 |
| | $1^1B_g$ | R | 6.45 | 6.42 | 1.00 | 6.22 |
| | $2^1A_u$ | R | 6.78 | 6.77 | 0.97 | 6.66 |
| | $2^1B_u$ | R | 7.23 | 7.41 | 0.89 | 7.07 |
| | $2^1B_g$ | R | 7.63 | 7.61 | 0.94 | 7.36 |
| | $3^1A_g$ | R | 7.78 | 7.78 | 0.94 | 7.62 |
| | $3^1B_u$ | R | 8.59 | 8.56 | 0.99 | 8.00 |
| pyridine | $^1B_1$ | V | 4.22 | 4.57 | 0.99 | 4.59 |
| | $^1B_2$ | V | 4.85 | 5.55 | 0.68 | 4.99 |
| | $^1A_2$ | V | 4.65 | 4.70 | 1.00 | 5.43 |
| | $^1A_1$ | V | 5.28 | 6.73 | 1.00 | 6.38 |
| pyrazine | $^1B_{3u}$ | V | 3.33 | 3.69 | 1.00 | 3.83 |
| | $^1B_{2u}$ | V | 4.44 | 5.47 | 0.80 | 4.81 |
| | $^1B_{2g}$ | V | 4.89 | 5.31 | 0.98 | 5.46 |
| | $^1B_{1g}$ | V | 5.79 | 5.86 | 1.00 | 6.10 |
| | $^1B_{1u}$ | V | 5.81 | 6.98 | 0.65 | 6.51 |
| pyridazine | $^1B_1$ | V | 2.94 | 3.33 | 0.99 | 3.60 |
| | $^1A_1$ | V | 4.91 | 5.65 | 0.64 | 5.00 |
| | $^1A_2$ | V | 4.88 | 5.16 | 0.98 | 5.30 |
| | $^1B_1$ | V | 5.53 | 5.63 | 0.99 | 6.00 |
| | $^1B_2$ | V | 5.56 | 5.82 | 1.00 | 6.50 |
| pyrimidine | $^1B_1$ | V | 3.72 | 3.94 | 0.99 | 3.85 |
| | $^1A_2$ | V | 4.13 | 4.22 | 0.98 | 4.62 |
| | $^1B_2$ | V | 5.04 | 5.80 | 0.67 | 5.12 |
| | $^1A_2$ | V | 4.95 | 5.26 | 0.97 | 5.52 |
| | $^1B_1$ | V | 5.36 | 5.55 | 0.99 | 5.90 |
| | $^1A_1$ | V | 5.45 | 6.99 | 0.66 | 6.70 |
| s-tetrazine | $^1B_{3u}$ | V | 1.58 | 1.93 | 1.00 | 2.25 |
| | $^1A_u$ | V | 2.93 | 3.06 | 0.99 | 3.40 |
| | $^1A_u$ | V | 4.34 | 4.66 | 0.98 | 5.00 |
| | $^1B_{3u}$ | V | 5.69 | 5.82 | 0.99 | 6.34 |
| Mean absolute error (total) | | | 0.40 | 0.27 | Average of $|c|^2$ : 0.93 | |
| Mean absolute error (valence) | | | 0.62 | 0.35 | | |
| Mean absolute error (Rydberg) | | | 0.23 | 0.21 | | |

## 5. Potential energy curves of thiophenol

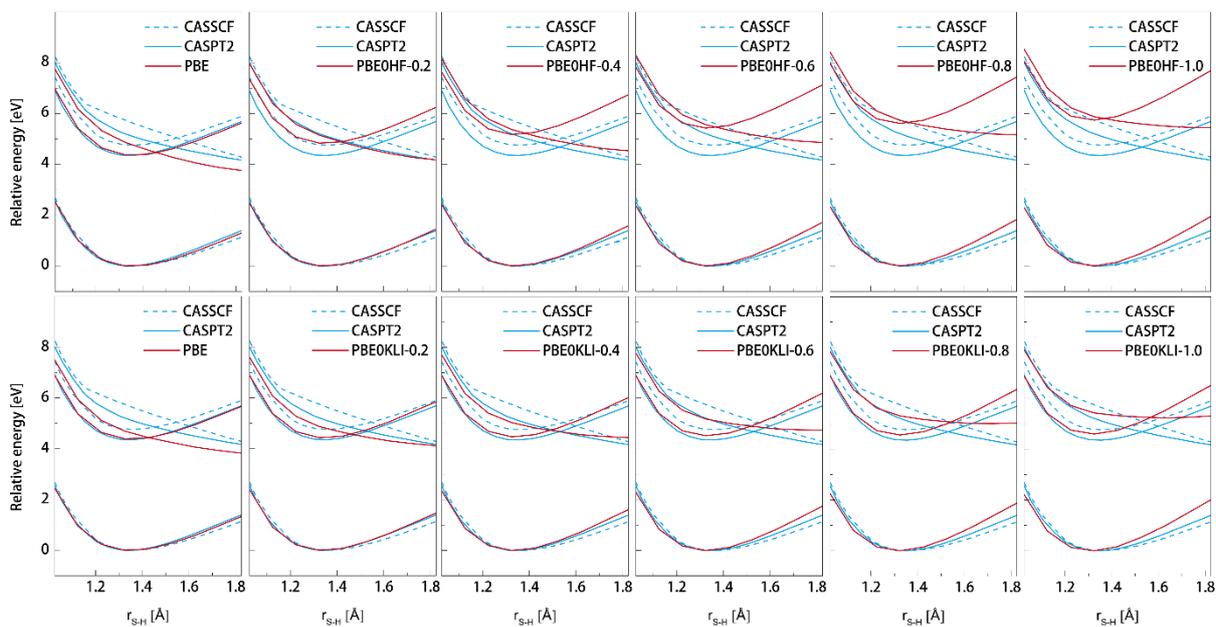

**Figure S2.** The potential energy curves of thiophenol obtained from TDDFT calculations. The ratio of the exact exchange is changed from 0 to 1 by 0.2. The upper and lower panels are the results of the Hartree-Fock and the optimized effective potential hybrid functionals, respectively. For practical reasons, we used the Krieger-Li-Iafrate (KLI) approximation for ground state calculations, and the local density approximation kernel for excited state calculations.

## 3. Molecular orbitals of thiophenol

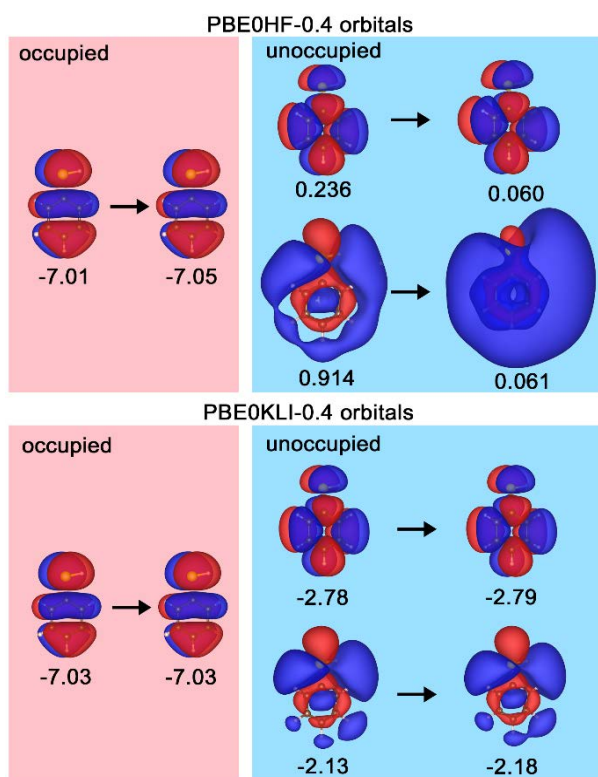

**Figure S3.** Comparison of occupied and virtual orbitals of thiophenol and their energies between the KLI and HF hybrid methods. For PBE0HF-0.4, the basis set increases from cc-pVTZ to aug-cc-pVTZ, while for PBE0KLI-0.4, the simulation box size increases from 12.0 Bohr to 18.0 Bohr. Unit: eV